\documentstyle[sprocl,epsfig]{article}

\def\be{\begin{equation}}
\def\ee{\end{equation}}
\def\bea{\begin{eqnarray}}
\def\eea{\end{eqnarray}}

\begin{document}

\title{\addvspace{-10mm} {\normalsize \hfill CFNUL/99-06 }\\
{\normalsize \hfill hep-ph/9912533 }\\
 \addvspace{5mm} 
RECENT DEVELOPMENT ON COLLECTIVE NEUTRINO INTERACTIONS
\footnote{Talk given at \emph{''Second
Meeting on New Worlds in Astroparticle Physics''}, Faro, Portugal, 1998.}
}
\author{LU\'{I}S BENTO}
\address{
Centro de F\'{\i}sica Nuclear da Universidade de Lisboa,\\
Av. Prof. Gama Pinto 2, 1649-003 Lisboa, Portugal }

\maketitle

%\begin{abstract}
\abstracts{Quantum Field Theory is applied to study an electron plasma under
an intense neutrino flux. The dispersion relation of the longitudinal waves
is derived and the damping rate is calculated. It is shown that in the case
of Supernova emission the neutrinos are not collimated enough to cause
plasma instabilities associated to a strong neutrino resonance effect. } 
%\end{abstract}

\section{Introduction}

Bingham {\em et.\ al.}\ \cite{bing94} have studied the behavior of an
electron plasma under a neutrino flux and concluded that neutrino fluxes as
intense as the ones produced in Supernovae cause plasma instabilities. If
true this could provide the physical mechanism of energy transfer from the
neutrinos to the medium that would explain the Supernovae explosion. The
weak interaction between the neutrinos and the plasma was described with the
concept of a ponderomotrive force acting on the electrons, 
\begin{equation}
\vec{F}_{{\rm pond}}=-\sqrt{2}\,G_{{\rm F}}\,c_{V}^{\prime }\,\vec{\nabla}%
n_{\nu }\ ,  \label{pond}
\end{equation}
proportional to the gradient of the neutrino number density ($c_{V}^{\prime
}=\pm 1/2+2\sin ^{2}\theta _{{\rm W}}$, ($+$) for $\nu _{e}$ and ($-$) for $%
\nu _{\mu }$, $\nu _{\tau }$). The neutrino wave function was assumed to
obey a naive Klein-Gordon equation modified to include matter effects.
However, that equation does not account for the neutrino spin degrees of
freedom and those works were subject to justified controversy \cite
{hard96,hard97}.

More recently, a classic kinetic theory was applied to describe the electron
and neutrino dynamics \cite{silv98}. In that work the neutrinos too are
subject to a ponderomotrive force analogous to Eq.\ (\ref{pond}) with $%
n_{\nu }$ replaced by the electron number density $n_{e}$. Again the lepton
spin and chiral structure of the weak interactions were ignored. In
addition, the dispersion relation derived for the plasma waves is
substantially different from the one previously obtained with the
Klein-Gordon equation \cite{bing94}. Nevertheless, the authors reiterated
the claim that the neutrinos produce resonant effects for certain modes of
plasma oscillation which cause instabilities characterized by large growth
rates proportional not to $G_{{\rm F}}^{2}\,n_{\nu }$ but to a smaller power
of that number.

The aim of the present work was to apply consistent Quantum Field Theory
techniques to obtain a kinetic description consistent with the quantum
mechanics and spin content of particles and interactions. First, we derive
the dispersion relation of the longitudinal waves of an electron plasma in
the presence of a neutrino flux and afterwards, analyze the possibility of
neutrino induced plasma instabilities.\strut

\section{Classic Kinetic Theory}

It proves useful to go first to a classic description before going to
Quantum Field Theory. The system of interest is a large number of electrons,
neutrinos and anti-neutrinos. Classically they are described by distribution
functions, $f_{e}$, $f_{\nu }$ and $f_{\bar{\nu}}$, on the respective
position and momentum phase spaces. In low dense plasmas the collisions are
less important than the collective interactions and the evolution of the
distribution functions $f(t,\vec{x},\vec{p})$ reduces to the Vlasov
equations \cite{lifs97,libo98}, 
\begin{equation}
\frac{\partial f}{\partial t}+\vec{v}\cdot \!\frac{\partial f}{\partial \vec{%
x}}+\vec{F}\!\cdot \!\frac{\partial f}{\partial \vec{p}}=0\ .  \label{Dfen}
\end{equation}
The dynamics is defined by specifying the velocity $\vec{v}$\ and force $%
\vec{F}$ as functions of $(t,\vec{x},\vec{p})$ for each of the particles
species. In the electron case $\vec{v}=\vec{p}/(\vec{p}^{2}+m_{e}^{2})^{1/2}$%
, but the neutrino mass will be neglected.

At low energies the electron interactions are described by the Standard
Model effective Lagrangian \cite{halz} 
\begin{equation}
{\cal L}_{{\rm int}}=e\,A_{\mu }\,\bar{e}\gamma ^{\mu }e-\sqrt{2}\,G_{{\rm F}%
}\,(\bar{\nu}_{L}\gamma _{\mu }\nu _{L})\,\bar{e}\,\gamma ^{\mu
}(c_{V}^{\prime }\,-c_{A}^{\prime }\,\gamma _{5})e\,\ ,  \label{Lint}
\end{equation}
where $e$ is the positron charge, $G_{{\rm F}}$ the Fermi constant, and $%
c_{V}^{\prime }=c_{A}^{\prime }+2\sin ^{2}\theta _{{\rm W}}$ with $%
c_{A}^{\prime }=+1/2$ for $\nu _{e}$ and $c_{A}^{\prime }=-1/2$ for $\nu
_{\mu }$, $\nu _{\tau }$. By analogy with electrodynamics one immediately
infers the Lagrangian of a classic {\em spinless} electron as 
\begin{equation}
L_{e}=\left( e\,A^{\mu }-\sqrt{2}\,G_{{\rm F}}\,c_{V}^{\prime }\,J_{\nu
}^{\mu }\,\,\right) \dot{x}_{\mu }\;.
\end{equation}
The average current density $J_{\nu }^{\mu }=\,\left\langle \bar{\nu}%
_{L}\gamma ^{\mu }\nu _{L}\right\rangle $, equal to the difference between
the neutrino and anti-neutrino current densities, $J_{\nu }^{\mu }=j_{\nu
}^{\mu }-j_{\bar{\nu}}^{\mu }$, acts in the same way as the electromagnetic
potential $A_{\mu }$. Hence, the electro-weak force applied to an electron
is a straightforward generalization of the Lorentz force, 
\begin{equation}
\vec{F}_{e}=-e(\vec{E}+\vec{v}_{e}\wedge \vec{B})+\sqrt{2}\,G_{{\rm F}%
}\,c_{V}^{\prime }\,(\vec{E}_{{\rm w}e}+\vec{v}_{e}\wedge \vec{B}_{{\rm w}%
e})\;\,,  \label{Fe}
\end{equation}
with weak-electric and weak-magnetic fields given by 
\begin{equation}
\vec{E}_{{\rm w}e}=-\vec{\nabla}J_{\nu }^{0}-\frac{\partial \vec{J}_{\nu }}{%
\partial t}\;,\qquad \vec{B}_{{\rm w}e}=\vec{\nabla}\wedge \vec{J}_{\nu }\;.
\label{EBwe}
\end{equation}
In a similar fashion, the weak forces applied to massless neutrinos is 
\begin{equation}
\vec{F}_{\nu }=\sqrt{2}\,G_{{\rm F}}\,(\vec{E}_{{\rm w}\nu }+\vec{v}_{\nu
}\wedge \vec{B}_{{\rm w}\nu })\   \label{Fn}
\end{equation}
for neutrinos ($\nu _{L}$) and $-\vec{F}_{\nu }$ for anti-neutrinos ($\bar{%
\nu}_{R}$), where 
\begin{equation}
\vec{E}_{{\rm w}\nu }=-\vec{\nabla}J_{{\rm w}e}^{0}-\frac{\partial \vec{J}_{%
{\rm w}e}}{\partial t}\;,\qquad \vec{B}_{{\rm w}\nu }=\vec{\nabla}\wedge 
\vec{J}_{{\rm w}e}\;,  \label{EBwn}
\end{equation}
are obtained from the weak electron current, 
\begin{equation}
J_{{\rm w}e}^{\mu }=\,\left\langle \bar{e}\,\gamma ^{\mu }(c_{V}^{\prime
}\,-c_{A}^{\prime }\,\gamma _{5})\,e\right\rangle =c_{V}^{\prime
}\,J_{e}^{\mu }-c_{A}^{\prime }\,J_{5e}^{\mu }\;.  \label{Jwe}
\end{equation}
It contains an axial current as well, that is different from zero if the
electrons are polarized: for an electron density $\,n_{e}$ and average
polarization $\left\langle \vec{\sigma}\right\rangle $, $\left\langle \,\bar{%
e}\,\vec{\gamma}\,\gamma _{5}\,e\right\rangle =\left\langle \vec{\sigma}%
\right\rangle n_{e}$.

The first point to note is that the weak forces contain other terms than the
ones identified before \cite{bing94,silv98} namely, those with the vector
current densities $\vec{J}_{\nu }$ and $\vec{J}_{{\rm w}e}$. That is true
even in the limit of non-relativistic and unpolarized electrons because the
neutrinos are ultra-relativistic.
Then, the following approximations apply: 
\begin{eqnarray}
\vec{F}_{e} &\cong &-e\,\vec{E}-\sqrt{2}\,G_{{\rm F}}c_{V}^{\prime }\,\left( 
\vec{\nabla}n_{\nu }-\vec{\nabla}n_{\bar{\nu}}+\frac{\partial \vec{j}_{\nu }%
}{\partial t}-\frac{\partial \vec{j}_{\bar{\nu}}}{\partial t}\right) \ ,
\label{Fenr} \\
\vec{F}_{\nu } &=&-\vec{F}_{\bar{\nu}}\;\cong \,-\sqrt{2}\,G_{{\rm F}%
}c_{V}^{\prime }\, \left( \vec{\nabla}n_{e} +
\frac{\partial \vec{j}_{e}}{\partial t} \right)  \ .  \label{Fnnr}
\end{eqnarray}
The electron current can be neglected, as presented in the conference, 
except for extreme wavelenghts, comparable to the inverse
plasma frequency times the velocity of light. 
For longitudinal waves $\vec{\nabla}\wedge \vec{j}_{e}$ is still zero and
does not contribute to the neutrino force.

The proper modes of a system are usually investigated by expanding the
distribution functions around certain zero order functions $f^{0}$. If the
background is static and uniform, $f^{0}$ only depend on the momenta and 
\begin{equation}
f(t,\vec{x},\vec{p})=f^{0}(\vec{p})+\delta f(t,\vec{x},\vec{p})\;.
\label{df}
\end{equation}
The Vlasov equations (\ref{Dfen}) are linearized by neglecting the quadratic
terms in $\delta f$ and the Fourier analysis is applied afterwards. Whenever
the forces vanish at zero order, as in the case of the electro-weak forces,
the Fourier transforms obey the equations 
\[
( \omega -\vec{k}\!\cdot \!\vec{v} ) \delta f(k,\vec{p})+\!i\vec{F}%
(k,\vec{p})\cdot \frac{\partial f^{0}}{\partial \vec{p}}=0\;,\, 
\]
where $\omega =k^{0}$ is the plasma frequency and $\vec{k}$ the wave vector.
The problem of deriving the dispersion relation is facilitated by the fact
that the forces (\ref{Fenr}), (\ref{Fnnr}) depend linearly on the current
densities 
\begin{equation}
j^{\mu }(t,\vec{x})=\int \!d^{3}\!p\,f(t,\vec{x},\vec{p})\,v^{\mu }\ 
\label{jmu}
\end{equation}
($v^{0}=1$). Then, the fluctuations 
\begin{equation}
\delta j^{\mu }(k)=\int \!d^{3}p\,\frac{-1}{\omega -\vec{k}\!\cdot \!\vec{v}}%
\frac{\partial f^{0}}{\partial \vec{p}}\!\cdot \!i\vec{F}(k,\vec{p})\,v^{\mu
}(\vec{p})\   \label{djmu}
\end{equation}
form a closed system of linear equations whose secular equation determines
the dispersion relation.

The Langmuir waves \cite{lifs97,libo98} are electron density waves
associated with an electrostatic field ($\vec{B}=\vec{0},\;{\rm rot\,}\vec{E}%
=\vec{0}$) whose potentials are 
\begin{equation}
A^{0}(k)=\frac{-e}{\vec{k}^{2}}\,\delta n_{e}(k)\ ,\qquad \vec{A}(k)=\vec{0}%
\;.  \label{A0}
\end{equation}
They are also called longitudinal electromagnetic waves and sometimes
plasmons in the Particle Physics literature \cite{bailin}. When the
electro-weak forces (\ref{Fenr}), (\ref{Fnnr}) are substituted in\ (\ref
{djmu}), one obtains $\delta n_{e}$
as a linear function of $\delta n_{e}$, $\delta n_{\nu }-\,\delta n_{\bar{\nu%
}}$, and $\delta \vec{j}_{\nu }-\,\delta \vec{j}_{\bar{\nu}}$ and, in turn, $%
\delta n_{\nu }$, $\,\delta \vec{j}_{\nu }$, $\,\delta n_{\bar{\nu}}$, $%
\,\delta \vec{j}_{\bar{\nu}}$ proportional to $\delta n_{e}$
($\delta \vec{j}_e = \omega \vec{k} \delta n_e  \! /\vec{k}^2$). 
It yields an
equation of the type $\delta n_{e}=-\chi \,\delta n_{e}$, where the
susceptibility $\chi $ has, in addition to the purely electromagnetic $\chi
_{_{EM}}$, a weak interaction term $\chi _{_{W}}$. The plasma dispersion
relation is 
\begin{equation}
1+\chi _{_{EM}}(\omega ,\vec{k})+\chi _{_{W}}(\omega ,\vec{k})=0\ .
\label{dr}
\end{equation}
$\chi _{_{EM}}$ only depends on the electron charge and distribution
function. It is shown in many books \cite{lifs97,libo98} that ($e^{2}=4\pi
\,\alpha $) 
\begin{equation}
\chi _{_{EM}} =\frac{e^{2}}{\vec{k}^{2}}\int \!d^{3}p\,\frac{%
1}{\omega -\vec{k}\!\cdot \!\vec{v}_{e}}\frac{\partial f_{e}^{0}}{\partial 
\vec{p}}\!\cdot \!\vec{k}\ .  \label{xiem}
\end{equation}
In the presence of a neutrino beam, the weak correction is given by 
\begin{equation}
\chi _{_{W}} =-2G_{{\rm F}}^{2}\,c_{V}^{\prime \,2}\frac{%
\chi _{_{EM}}}{e^{2}} \left( 1- \frac{\omega^2}{\vec{k}^2} \right)
\int \!d^{3}p\,\frac{\vec{k}\!\cdot \!(\vec{k}-\omega \,%
\vec{v}_{\nu })}{\omega -\vec{k}\!\cdot \!\vec{v}_{\nu }}\left( \frac{%
\partial f_{\nu }^{0}}{\partial \vec{p}}+\frac{\partial f_{\bar{\nu}}^{0}}{%
\partial \vec{p}}\right) \!\cdot \!\vec{k}\ .  \label{xiw}
\end{equation}
Notice that despite the relative sign in the expression of the electron
force, Eq.\thinspace (\ref{Fenr}), the $\nu $ and $\bar{\nu}$ beams
contribute in the same way to the plasma susceptibility. The reason is the $%
\nu $ and $\bar{\nu}$ density fluctuations get also opposite signs from the
forces $\vec{F}_{\nu }$ and $\vec{F}_{\bar{\nu}}$. That expression of $\chi
_{_{W}}$ differs from the result obtained in\ \cite{silv98} by the term $%
-\omega \,\vec{v}_{\nu }$ under the integral and the factor
$1-\omega^2 \!/ \vec{k}^2$, which come from the time
derivative of $\vec{J}_{\nu }$ and $\vec{j}_{e}$ in the forces (\ref{Fenr}) and
(\ref{Fnnr}) respectively.

%\newpage

\section{Quantum Field Theory}

A basic property of the weak interactions is their chiral structure {\em i.e.%
}, the differentiation between the left and right-handed components of the
fermions. This aspect is ignored in the classic theory because it does not
include the dynamics of the spin degrees of freedom. At the fundamental
level the basic entities are the quantum fields but in the situation of
interest they interact within a bath of plasma and neutrinos. For a medium
in thermal equilibrium there are well established Finite Temperature Field
Theory techniques \cite{bailin} however, the passing neutrinos are not in
thermal equilibrium and the plasma waves represent themselves deviations
from equilibrium. \strut The matter content is represented by a density of
states $\rho $. In the Heisenberg picture, $\rho $ is constant and the
fields evolve in time. In the Standard Model they are subject to local,
relatively simple interactions and we can take advantage of that to solve
the field equations of motion perturbatively \cite{bent98}.

In matter, the one-particle propagators are different from the propagators
in vacuum. I define the distribution functions as a difference between the
expectation values in matter and in vacuum of the time-order products of the
fields in the Heisenberg picture {\em i.e.}, 
\begin{equation}
\left\langle T\,\psi ^{\alpha }(x)\,\bar{\psi}^{\beta }(y)\right\rangle
_{\rho }=\left\langle 0\left| T\,\psi ^{\alpha }(x)\,\bar{\psi}^{\beta
}(y)\right| 0\right\rangle -f^{\alpha \beta }(x,y)\ .  \label{fab}
\end{equation}
That $f^{\alpha \beta }(x,y)$\ represent one-particle distribution functions
can been seen by evaluating observables like the current density 
\begin{eqnarray}
J^{\mu }(x) &=&\left\langle \bar{\psi}(x)\,\gamma ^{\mu }\,\psi
(x)\right\rangle _{\rho }-\left\langle 0\left| \bar{\psi}(x)\,\gamma ^{\mu
}\,\psi (x)\right| 0\right\rangle  \label{Jmuq} \\
&=&{\rm tr}\left\{ f(x,x)\,\gamma ^{\mu }\right\} \ ,  \nonumber
\end{eqnarray}
where the trace is over the spinor indices.

In absence of interactions one knows the exact expressions of the quantum
fields and exact distribution functions can be easily specified. The
function 
\begin{equation}
f_{e}^{0}(x,y)=\int \!\frac{d^{3}p}{4\,E_{e}}\,f_{e}^{0}(\vec{p})\,\,(p%
\hspace{-0.19cm}/_{e}+m_{e})\,e^{-i\,p_{e}\cdot (x-y)}\   \label{fe0}
\end{equation}
describes a homogeneous (invariant under translations) distribution of
unpolarized electrons with momentum distribution $f_{e}^{0}(\vec{p})$, as
shows the expression of the current density, 
\[
J_{e}^{\mu }{}^{0}(x)={\rm tr}\left\{ f_{e}^{0}(x,x)\,\gamma ^{\mu }\right\}
=\int \!d^{3}p\,f_{e}^{0}(\vec{p})\,\frac{p_{e}^{\mu }}{E_{e}}\ . 
\]
Likewise, uniform beams of {\em massless}\ neutrinos have a distribution
function (associated to the field $\nu _{L}$) 
\begin{equation}
f_{\nu }^{0}(x,y)=\frac{1-\gamma _{5}}{2}\int \!\frac{d^{3}p}{2E_{\nu }}%
\,\left( f_{\nu }^{0}(\vec{p})\,p\hspace{-0.19cm}/_{\nu }\,e^{-i\,p_{\nu
}\cdot (x-y)}-f_{\bar{\nu}}^{0}(\vec{p})\,p\hspace{-0.19cm}/_{\nu
}\,e^{+i\,p_{\nu }\cdot (x-y)}\right) \ ,  \label{fn0}
\end{equation}
where $f_{\nu }^{0}$ and $f_{\bar{\nu}}^{0}$ represent the (left-handed)
neutrino and (right-handed) anti-neutrino momentum distributions. $%
f_{e}^{0}(x,y)$ and $f_{\nu }^{0}(x,y)$ will be used as background
distribution functions which implies a plasma with no positrons and
therefore temperatures below the electron mass.

%\vspace{3mm}
\begin{figure}[b]
\centering
\epsfig{figure=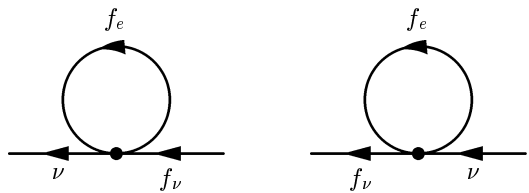,width=7cm}
\caption{Diagrams contributing to $\delta f_\nu$.}
\end{figure}

The functions $f_{e}^{0}(x,y)$ and $f_{\nu }^{0}(x,y)$ stop to be consistent
with the definition (\ref{fab}) when the field equations of motion include
interactions. The deviations 
\begin{equation}
\delta f(x,y)=f(x,y)-f^{0}(x,y)
\end{equation}
can be calculated perturbatively \cite{bent98} as part of higher order
corrections to the one-particle Green functions (\ref{fab}) along with other
radiative corrections such as vacuum self-energies and matter induced
'potential' energies. $\delta f_{e}$ and $\delta f_{\nu }$ are the
corrections associated to the very matter fluctuations $\delta f(x,y)$ and,
in first approximation, they are given by the diagrams in Figs. 1 and 2.

At one-loop, the Fourier transforms of $\delta f(x,x)$, $\delta f_{e}(k)$
and $\delta f_{\nu }(k)$, are just linear combinations of the leptonic
current fluctuations namely, $\delta J_{\nu }^{\mu }={\rm tr}\left\{ \delta
f_{\nu }\gamma ^{\mu }\right\} $, $\delta J_{e}^{\mu }={\rm tr}\left\{
\delta f_{e}\gamma ^{\mu }\right\} $ and $\delta J_{5e}^{\mu }={\rm tr}%
\left\{ \delta f_{e}\gamma ^{\mu }\gamma _{5}\right\} $. As a result, $%
\delta J_{e}^{\mu }(k)$, $\delta J_{5e}^{\mu }(k)$ and $\delta J_{\nu }^{\mu
}(k)$ form a closed system of coupled equations, $\,\delta J=A\,\delta J$ in
a matrix notation. In principle one could derive the dispersion relation $%
\omega (\vec{k})$ from the condition $\det (1-A)=0$ but that is not easy. In
particular, the Lorentz structure is non-trivial because the plasma rest
frame, wave vector $\vec{k}$ and neutrino fluxes define privileged
directions in space-time. The first possible simplification is to take the
macroscopic limit by neglecting $\omega $ and $\vec{k}$\ with respect to the
one-particle energies $E_{e}$, $E_{\nu }$, making in particular $(p_{e}\pm
k)^{2}-m_{e}^{2}\approx \pm 2k\cdot p_{e}$ and $(p_{\nu }\pm k)^{2}\approx
\pm 2k\cdot p_{\nu }$ in the electron and neutrino propagators. It turns out
that in this limit the field theory results for the currents $\delta
J_{e}^{\mu }$ and $\delta J_{\nu }^{\mu }$ are equivalent to the classic
expressions (\ref{djmu}) of $\delta j_{e}^{\mu }$ and $\delta j_{\nu }^{\mu
}-\delta j_{\bar{\nu}}^{\mu }$ respectively, with forces $\vec{F}_{e}(k,\vec{%
p})$\ and $\vec{F}_{\nu }(k,\vec{p})$\ given exactly by the Fourier
transforms of the forces (\ref{Fe}) and (\ref{Fn}), derived from the
appropriate classic theory.

%\vspace{7mm}

\begin{figure}[t]
\centering
\epsfig{figure=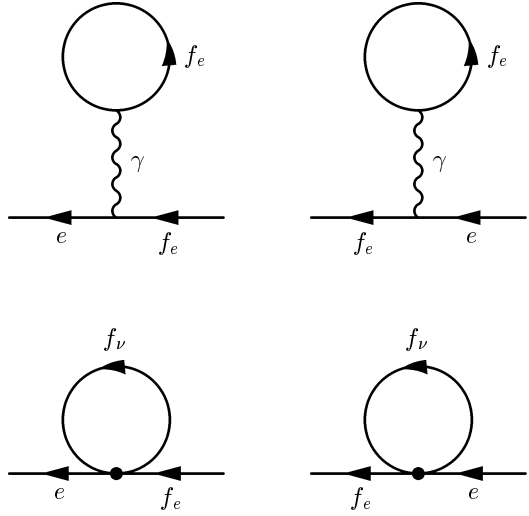,width=7cm}
\caption{Diagrams contributing to $\delta f_e$.}
\end{figure}

 Of course, the axial current $\delta J_{5e}^{\mu }$ can only be calculated
within field theory. One expects from the Standard Model Lagrangian 
(\ref{Lint}) that the neutrino vector current polarizes the electron spin like 
a magnetic field does on a magnetic moment. 
The density $\delta J_{5e}^{0}$ is suppressed for non-relativistic electrons
but the axial current $\delta \vec{J}_{5e}$ gets polarized along the 
component of the neutrino velocity orthogonal to the wavevector $\vec{k}$. 
It is also accompained by a transverse electric current both of them 
proportional to $G_{{\rm F}}^{2}$. 
However, their contribution to the dispersion relation only comes at the $G_{{\rm F}}^{4}$ order and can so be discarded. 
The classic dispersion relation (\ref{xiw}) is recovered for the 
non-relativistic plasma.

\section{Plasma oscillation. Damping.}

Let $\omega _{{\rm pl}}(\vec{k})$ denote the plasma frequency as a function
of $\vec{k}$ in the absence of neutrinos. For wavelengths much larger than
the Debye radius \cite{lifs97,libo98}, the electromagnetic susceptibility is
given by $\chi _{_{EM}}(\omega ,\vec{k})=-\omega _{{\rm pl}}^{2}(\vec{k}%
)/\omega ^{2}$. After applying this relation in the Eqs. (\ref{dr}) and (\ref
{xiw}) the dispersion relation\ of the non-relativistic plasma in the
presence of neutrinos reads as 
\begin{equation}
\omega ^{2}-\omega _{{\rm pl}}^{2}(\vec{k})=-\frac{\,G_{{\rm F}%
}^{2}\,c_{V}^{\prime \,2}}{2\pi \alpha }
\frac{\omega _{{\rm pl}}^{2}(\vec{k})}{\vec{k}^{2}} 
(\vec{k}^{2}-\omega ^{2}) \!
\int \! \!
d^{3}p\,\frac{\vec{k}\!\cdot \!(\vec{k}-\omega \,\vec{v}_{\nu })}{\omega -%
\vec{k}\!\cdot \!\vec{v}_{\nu }}\left( \frac{\partial f_{\nu }^{0}}{\partial 
\vec{p}}+\frac{\partial f_{\bar{\nu}}^{0}}{\partial \vec{p}}\right) \!\cdot
\!\vec{k}\ ,  \label{w1}
\end{equation}
or, after integrating by parts, 
\begin{equation}
\omega ^{2}-\omega _{{\rm pl}}^{2}(\vec{k})=\frac{\,G_{{\rm F}%
}^{2}\,c_{V}^{\prime \,2}}{2\pi \alpha }
\frac{\omega _{{\rm pl}}^{2}(\vec{k})}{\vec{k}^{2}} 
(\vec{k}^{2}-\omega ^{2})^2
\int \!\frac{d^{3}p}{E_{\nu }}\left( f_{\nu }^{0}+f_{\bar{%
\nu}}^{0}\right) \,\frac{\vec{k}^{2}\!-(\vec{k}\!\cdot \!\vec{v}_{\nu })^{2}%
}{ ( \omega -\vec{k}\!\cdot \!\vec{v}_{\nu } ) ^{2}}\ .  \label{w2}
\end{equation}
Again, as a consequence of the discrepancies in the weak forces, these
results differ from \cite{silv98} namely, by the factor 
$(1-\omega ^{2} \!/\vec{k}^{2})^2$ in the second equation.

The constant $c_{V}^{\prime \,2}$ is much smaller for $\nu _{\mu }$, $\nu
_{\tau }$ ($c_{V}^{\prime }\approx -0.04$) than for $\nu _{e}$ ($%
c_{V}^{\prime }\approx 0.96$), so only $\nu _{e}$ and $\bar{\nu}_{e}$
deserve to be taken in consideration. The expressions above provide a good
perspective of the real impact of the neutrinos. Keeping only the main
factors, 
\[
\omega ^{2}-\omega _{{\rm pl}}^{2}\;\propto \;\omega _{{\rm pl}}^{2}\,\frac{%
G_{{\rm F}}\,n_{\nu }}{E_{\nu }}\,G_{{\rm F}}\,\vec{k}^{2}\; 
\]
clearly indicates that the neutrino effect is severely suppressed by $G_{%
{\rm F}}^{2}$. The only potential exception are the oscillation modes that
catch the neutrinos in the resonance $\omega =\vec{k}\!\cdot \!\vec{v}_{\nu
} $.

If the neutrino spectrum goes from one side of the pole to the other the
situation is analogous to the electron Landau damping \cite{lifs97,libo98}.
The integral in Eq.\ (\ref{w1}) separates into a negligible principal part
and a complex quantity that is determined by the so-called Landau
prescription, here $\omega -\vec{k}\!\cdot \!\vec{v}_{\nu }+i\varepsilon
,\;\varepsilon \rightarrow 0^{+}$ (identifies with the Feynman boundary
conditions on the propagators). Denoting the imaginary part of $\omega $ as $%
-i\gamma $, the neutrino contribution to the decay rate $\gamma $ is then 
\begin{equation}
\gamma _{_{{\rm W}}}=-\frac{\,G_{{\rm F}}^{2}\,c_{V}^{\prime \,2}}{4\alpha }%
\, \frac{\omega}{\vec{k}^{2}} \,
(\vec{k}^{2}-\omega ^{2})^2 \!\int \!d^{3}p\,\,\delta (\omega -\vec{k%
}\!\cdot \!\vec{v}_{\nu })\left( \frac{\partial f_{\nu }^{0}}{\partial \vec{p%
}}+\frac{\partial f_{\bar{\nu}}^{0}}{\partial \vec{p}}\right) \!\cdot \!\vec{%
k}\ .  \label{gweak}
\end{equation}
Hardy and Melrose \cite{hard97} obtained this result from the calculation 
of the decay rate of neutrinos into longitudinal photons. However, 
by the very nature of such calculation the full neutrino contribution to the 
dispersion relation  is not derived and second, cannot be utilized to 
investigate possible hydrodynamic instabilities.

$\gamma _{_{{\rm W}}}$ goes as $G_{{\rm F}}^{2}$ and is exceedingly small.
If however all the neutrinos were on the top of the resonance they could
generate an hydrodynamic instability. The claim \cite{bing94,silv98} was
that this occurs in the conditions of Supernova neutrino emission causing
much larger growth rates. The question I rise is, in the end one has to
check whether or not the entire neutrino flux lies in the resonance {\em i.e.%
}, whether $|\omega -\vec{k}\!\cdot \!\vec{v}_{\nu }|$ stays within the
calculated resonance width $|\gamma |$. The neutrino velocities only spread
in direction. At radii, $r$, much larger than the neutrinosphere radius, $%
R_{\nu }$, they move essentially in the radial direction yet, the angular
spread $\theta _{\nu }\approx 2R_{\nu }/r$ is finite. Under the resonance
condition, $\vec{k}\!\cdot \!\vec{v}_{\nu }\approx \omega _{{\rm pl}}$, the
interval of variation of $\vec{k}\!\cdot \!\vec{v}_{\nu }=|\vec{k}|\,\cos
\theta _{\nu k}$ lies in the range 
\begin{equation}
\theta _{\nu }^{2}\,\omega _{{\rm pl}}\leq \,\Delta \vec{k}\!\cdot \!\vec{v}%
_{\nu }\approx \theta _{\nu }\tan \theta _{\nu k}\,\omega _{{\rm pl}}\,\leq
\omega _{{\rm pl}}\;.
\end{equation}
The left-hand value holds for a vector $\vec{k}\!$ with radial direction
whereas the right-hand bound is necessary to exclude the situation of Landau
damping and the result (\ref{gweak}).

The largest values of $|\gamma |$ are obtained by assuming that $\vec{k}%
\cdot \vec{v}_{\nu }$ is approximately constant over the neutrino spectrum
and can be factorized out of the integral in Eq.\ (\ref{w2}). One obtains
growth rates of the order of 
\[
\gamma _{{\rm optimistic}}=\omega _{{\rm pl}}\left\{ \frac{G_{{\rm F}%
}\,n_{\nu }}{E_{\nu }}\frac{G_{{\rm F}}\,n_{e}}{E_{e}}\right\} ^{1/3}\left( 
\frac{\vec{k}^{2}-\omega _{{\rm pl}}^{2}}{\omega _{{\rm pl}}^{2}}\right)
^{2/3}\; 
\]
for the resonant modes ($\vec{k}\!\cdot \!\vec{v}_{\nu }\approx \omega _{%
{\rm pl}}$). $\gamma _{{\rm opt.}}$ varies with $\tan ^{4/3}\theta _{\nu k}$
and $\Delta \vec{k}\!\cdot \!\vec{v}_{\nu }$ with $\tan \theta _{\nu k}$, so
the most favorable case for a hydrodynamic instability is a vector $\vec{k}$
orthogonal to the radial direction i.e., $\Delta \vec{k}\!\cdot \!\vec{v}%
_{\nu }\approx \omega _{{\rm pl}}$\ and $\tan \theta _{\nu k}\approx
1/\theta _{\nu }$. Then, the condition that the neutrinos be inside the
resonance requires 
\[
\hat{\gamma}=\left\{ \frac{G_{{\rm F}}\,n_{\nu }}{E_{\nu }}\frac{G_{{\rm F}%
}\,n_{e}}{E_{e}}\right\} ^{1/3}\theta _{\nu }^{-4/3}\;\geq \frac{\gamma _{%
{\rm optimistic}}}{\Delta \vec{k}\!\cdot \!\vec{v}_{\nu }}\geq \;1\;. 
\]
Knowing how small the energies $G_{{\rm F}}\,n_{\nu }$ and $G_{{\rm F}%
}\,n_{e}$ are, that looks quite unlikely. For numerical estimates I used the
values \cite{mayl87} $L_{\nu _{e}}=10^{53}\,{\rm ergs/s}$, $E_{\nu }=10\,%
{\rm MeV}$, $n_{e}=10^{33}\,{\rm cm}^{-3}$ and $\theta _{\nu }=100\,{\rm km}%
/r$. For $\theta _{\nu }=0.1$, $\hat{\gamma}$ is of the order of $10^{-7}$,
quite small indeed. Furthermore, since the neutrino density varies as $%
1/r^{2}$, $\gamma _{{\rm opt.}}/ \omega _{{\rm pl}}$ is proportional to $%
(n_{e}r^{2})^{1/3}$ and necessarily drops to zero for large enough radius.
One may also fix the value of the electron density. It turns out that the
parameter $\hat{\gamma}$ approaches the unity only at a radius equal to $%
10^{13}\,{\rm km}$ for $n_{e}$ as large as $10^{33}\,{\rm cm}^{-3}$! That is
clearly absurd.\strut

It should be also taken in consideration that for wavelengths smaller than
the Debye distance ($k>k_{{\rm D}}$), the plasma waves are
electromagnetically Landau damped \cite{lifs97,libo98}. That puts a further
upper limit on the interesting values of $\tan \theta _{\nu k}$ 
for the resonant neutrinos ($<k_{D}/$ $%
\omega _{{\rm pl}}$) in disfavour of the
hydrodynamic instabilities.

\section{Conclusions}

We have shown that Quantum Field Theory yields results that are compatible
with the classic kinetic theory provided that one uses the right
specification for the forces. Simple considerations show that the weak
forces between electrons and neutrinos are of the same type as the Lorentz
electromagnetic force. In particular, the force done by a neutrino flux on a
non-relativistic electron is 
\begin{equation}
\vec{F}_{e}=-\sqrt{2}\,G_{{\rm F}}\,c_{V}^{\prime }\,\left( \vec{\nabla}%
n_{\nu }+\frac{\partial \vec{j}_{\nu }}{\partial t}\right) \ .
\end{equation}
In Quantum Field Theory the spin degrees of freedom are also taken in
account. Under a neutrino flux the electrons get polarized at $G_{{\rm F}%
}^{2}$ order but the final effect on the plasma dispersion relation is of
higher order and can be neglected.

In what concerns possible plasma instabilities caused by the neutrino flux
in a Supernova envelope star we have shown that the neutrinos are never
collimated enough to produce strong resonance effects. As a consequence, the
neutrino induced growth rates are of the Landau damping type,
suppressed by $G_{{\rm F}}^{2}$.

\section*{Acknowledgements}

I thank Tito Mendon\c{c}a, Lu\'{\i }s Silva and Ana Mour\~{a}o for valuable
discussions. This work was supported in part by FCT under the grants
ESO/P/PRO/ 1127/96 and PESO/P/PRO/1250/98.

\end{document}